\documentclass[11pt]{article}
\usepackage{moriond,graphicx}

\def\Journal#1#2#3#4{{#1} {\bf #2}, #3 (#4)}

\def\NPB{{\em Nucl. Phys.} B}
\def\PLB{{\em Phys. Lett.}  B}

\def\PR{\em Phys. Rep.}
\def\PRD{{\em Phys. Rev.} D}

\def\be{\begin{equation}}
\def\ee{\end{equation}}
\def\bea{\begin{eqnarray}}
\def\eea{\end{eqnarray}}

\begin{document}
\vspace*{4cm}
\title{Recent Tevatron Searches}

\author{ J.T. Linnemann}
\author{ (For the D\O\ and CDF Collaborations)}

\address{linnemann@pa.msu.edu\\Department of Physics \& Astronomy, BPS Building,\\
E. Lansing, MI 48824, USA}

\maketitle\abstracts{ Here I present some recent results of the
D\O\  and CDF collaborations on Large Extra Dimensions,
$Z^\prime$, and SUSY searches.  The experiments examine events
produced by proton-antiproton collisions, with an integrated
luminosity of approximately 200 events/pb per experiment.
Unfortunately, despite careful searches, no new signals of
physics beyond the standard model were observed. Therefore, I
present the limits derived from the observations.}

\section{Introduction}
The analyses selected here are for the most part new results
since the 2003 Lepton-Photon conference.  These are all analyses
by the D\O\ ~\cite{d0page} and CDF~\cite{cdfpage} collaborations
of Run II data taken since 2002, with the upgraded detectors and
the raised center of mass energy of 1.96 TeV. Given the
constraints on time and space, not even every new result will be
covered.  The limits shown are
 95\% confidence limits unless otherwise stated, but
these limits are derived with prescriptions which vary from
analysis to analysis.
\section{Large Extra Dimensions (LED)}
Both experiments have new results on searches for the effects of
Large Extra Dimensions (compactified spatial dimensions beyond
the 3 we are used to).  It is instructive to contrast the
strategies used in these searches. D\O\ looks in both the ee and
e$\gamma$ channels, while CDF uses only the ee channel.  Both
emphasize events with at least one of the particles in their
central rapidity region ($|\eta| <\ $1 for CDF; or 1.1 for D\O )
because of large QCD backgrounds for cases where both particles
are at extreme rapidities ($1.1 < |\eta| < 2.8$ for CDF; 1.5 to
2.4 for D\O\ ). Both study the 2-particle mass spectra for
deviations from the standard model. However, here things
diverge.  CDF bases its limits primarily on the mass spectrum,
and calculates generic efficiencies based on the spin of the
2-particle system, which (with the mass) governs the angular
distribution.  The result is a limit (see Fig 1a) on $\sigma
\cdot B$ (cross section times branching ratio) for each spin.
This allows them to compare many models (and predicted cross
sections) with the mass spectrum with little extra model-specific
effort. D\O\ has chosen instead to aim at more optimized analyses
of specific models. For example, they fit the two-dimensional
distribution (Fig 1b) of mass and cos$\theta^*$ ( center of mass
angle) to a particular LED model plus  standard model and
instrumental backgrounds.  Further, the analysis is performed
separately for events with both, or only one, particle in the
central rapidity region.

\begin{figure}[ht]
  \begin{center}
  \begin{tabular}{cc}
 \includegraphics[width=77mm,height=2.2in]{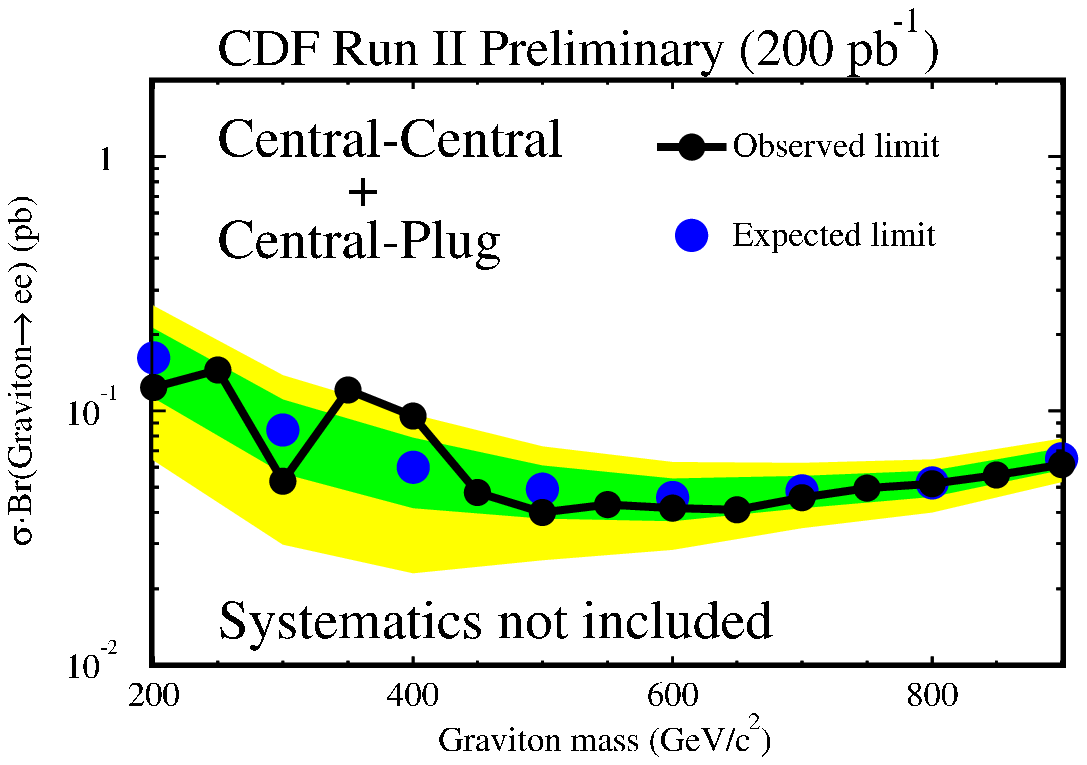} &
   \includegraphics[width=77mm,height=2.2in]{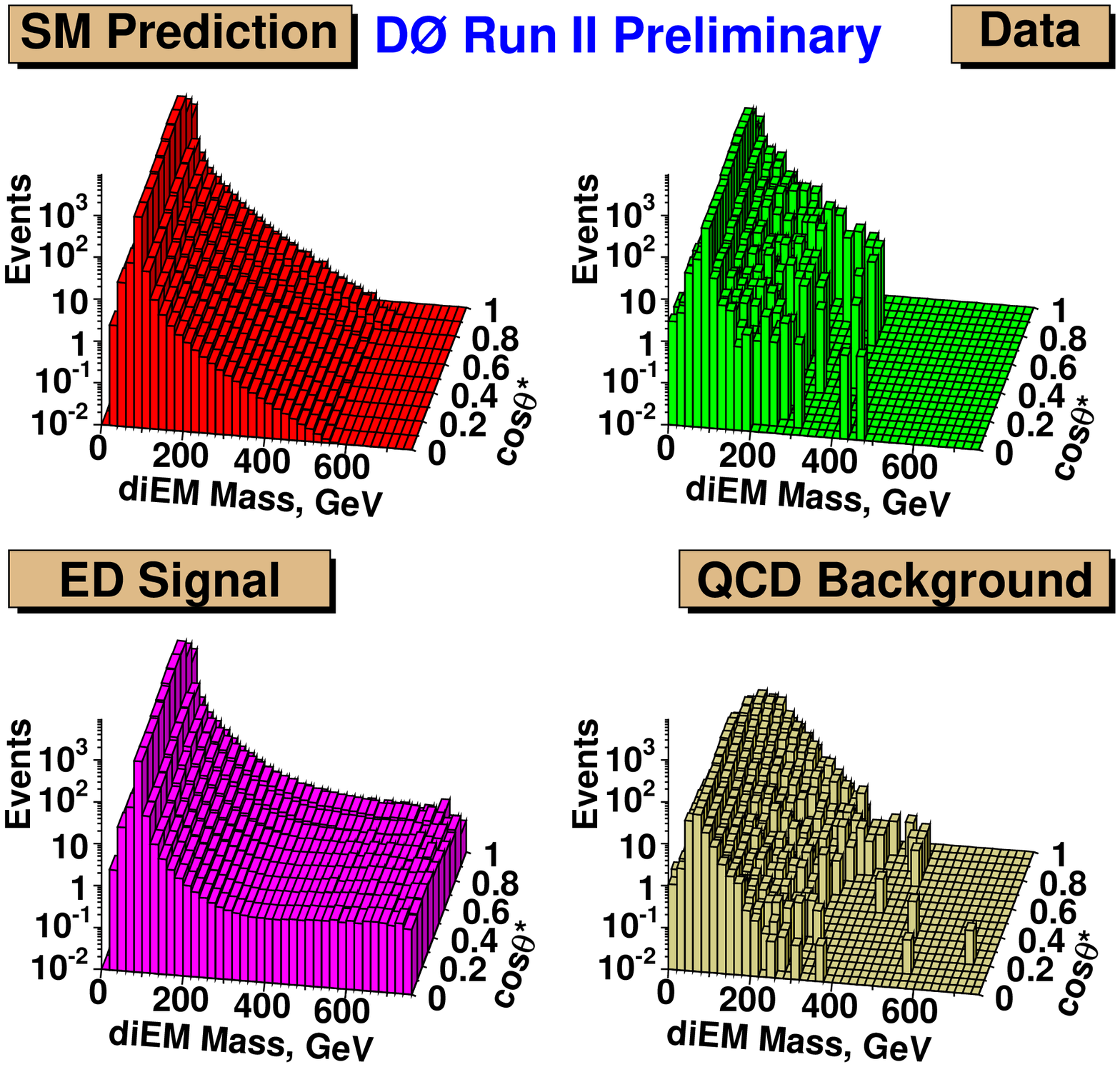}
  \end{tabular}
  \caption{Left (a): CDF expected and achieved limit ee data,
  assuming spin 2.  The inner and outer bands show 1 and 2
  standard deviations about the expected limit.  Final achieved
  limits (not shown) have systematics included.
   Right (b): D\O\ distributions for two electromagnetic
   showers as a function of mass and cos $\theta^*$ for
   background, data, and signal.}
  \label{fig:contrast}
  \end{center}
  \end{figure}

\subsection{ADD Extra Dimensions}
The first limit derived from these data is for the so-called
ADD~\cite{add} extra dimensions, for which SM particles are
confined to a D3-brane, while gravity propagates in the n extra
dimensions as well, explaining its apparent weakness. Both the CDF
and D\O\ analyses are based an integrated luminosity of 200
events/pb.  Figure 2 shows the results from each experiment as a
function of mass. The expected cross section contributions beyond
the standard model can be parameterized in terms of $\eta_{G} =
F/M^4_S$.  The limits on the cross section and thus on $\eta$ can
be interpreted in terms of $M_S$ (the 3+n dimensional Planck
scale) by use of the GRW~\cite{grw} convention, in which $F = 1$.
This coincides with the $HLZ$~\cite{hlz} convention for $F =
2/(n-2)$ when $n=4$. The result for the CDF Run II
analysis~\cite{cdfee} is $M_S
> 1.11\ $TeV while the more optimized D\O\ analysis~\cite{d0led} produces a
limit of $M_S > 1.36\ $ TeV. D\O\ also combines their Run I and
Run II results for a limit of $M_S > 1.43\ $TeV, the most
restrictive limit to date.

\begin{figure}[ht]
  \begin{center}
  \begin{tabular}{cc}
    \includegraphics[width=70mm,height=2.5in]{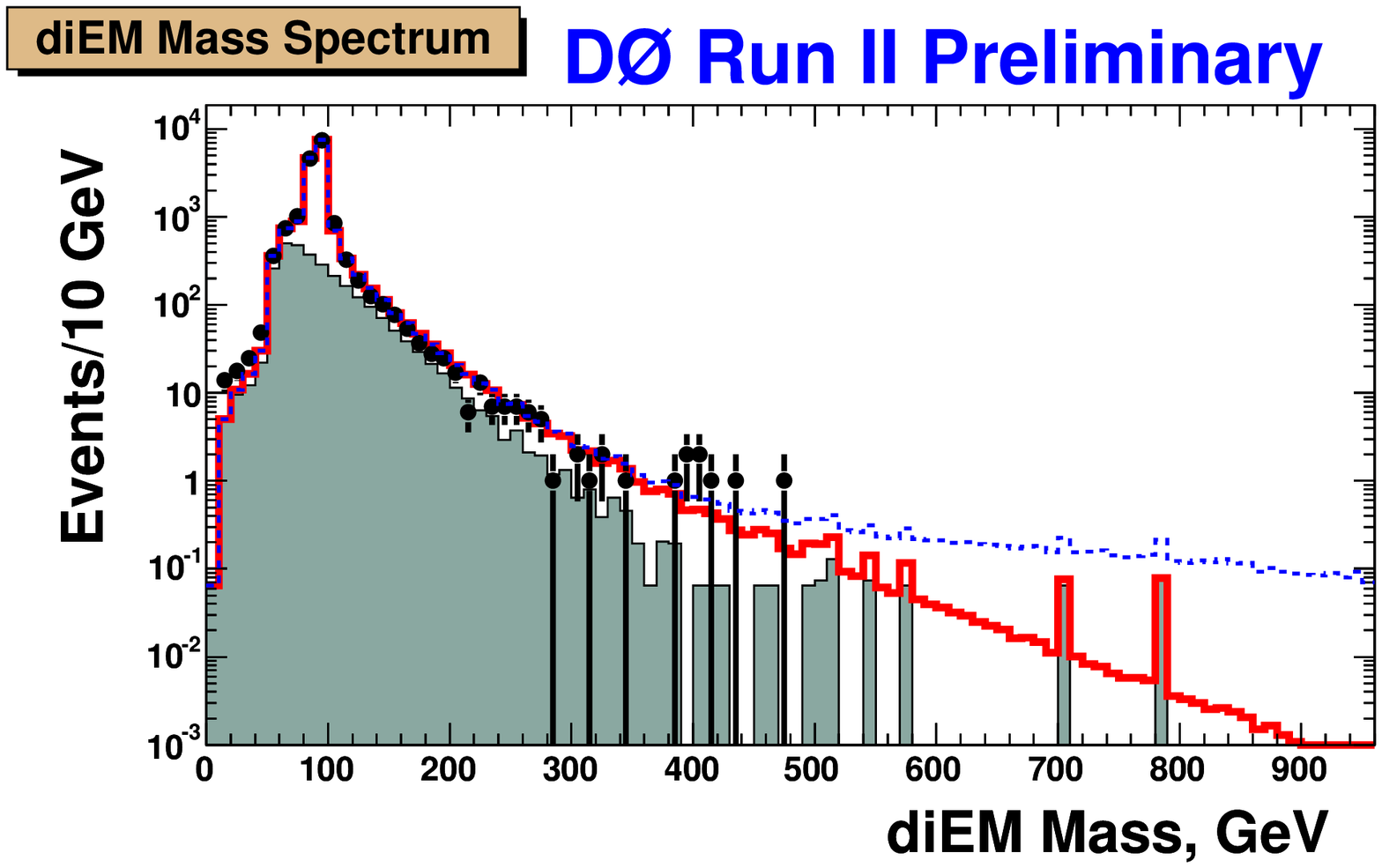}
    &      \includegraphics[width=84mm,height=2.5in]{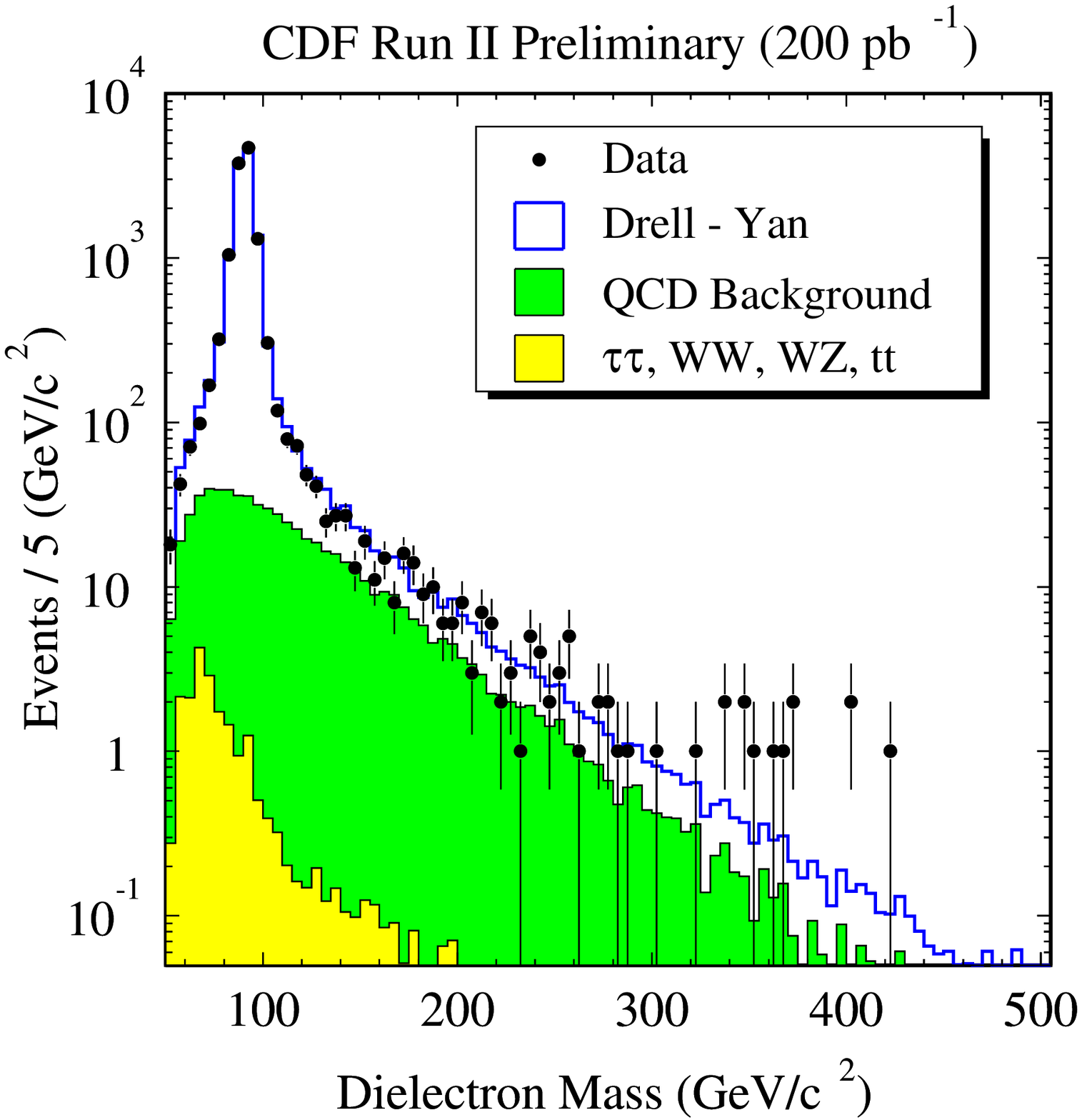}
  \end{tabular}
  \caption{Left: D\O\ mass spectrum for two electromagnetic objects.
  The shaded histogram indicates the instrumental
  background; the solid curve a fit to background + SM; the dashed
  curve adds a LED signal for $\eta_G = .6$.
  Right: CDF dielectron spectrum with SM expectations.}
  \label{fig:eespectra}
  \end{center}
  \end{figure}

For those prone to excitement, D\O\ has examined the
low-statistics ``bump'' in the mass spectrum at 400 GeV.  Few of
the events are consistent with being electrons, and the width is
well below the expected mass resolution.

\subsection{TeV$^{-1}$ (Longitudinal) Extra Dimensions}
An alternative realization~\cite{tevdim} of extra dimensions is
the TeV$^{-1}$ scheme, in which matter resides on a p-brane, with
chiral fermions on the 3-brane internal to the p-brane and SM
gauge bosons propagating in all p dimensions.  This scenario
results in a compatification scale of 1/$M_c$.  For dielectrons,
this is equivalent to Kaluza-Klein towers of gauge bosons of mass
$M_n = \sqrt{M^2 + n^2 M^2_c}$ (here M is either the Z or
$\gamma$). Unlike LED, this model gives negative interference
 at intermediate masses and an enhancement of the cross
section at higher masses, as seen in Fig 3a.  D\O\ has performed
a dedicated search~\cite{d0teved} in the ee final state. This is
the first direct search for the effects of these virtual KK
exchanges, and it produces a limit of $M_c > 1.12$ TeV. Indirect
searches~/cite{lands} at LEP imply $M_c
> 6.6$ TeV.

\begin{figure}[ht]
  \begin{center}
  \begin{tabular}{cc}
  \includegraphics[width=77mm]{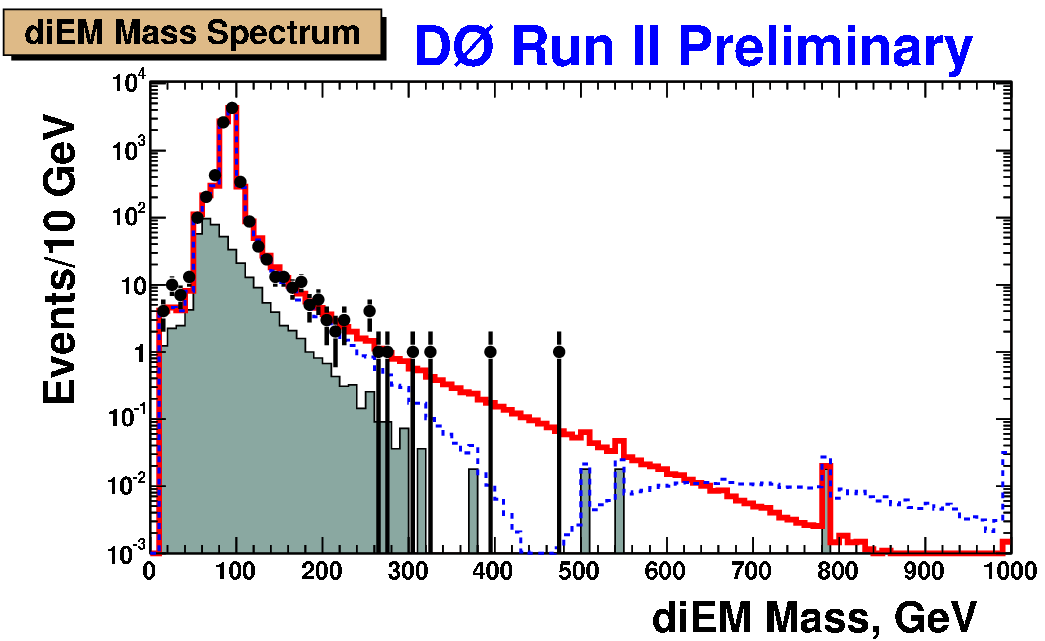} &
    \includegraphics[width=77mm,height=1.9in]{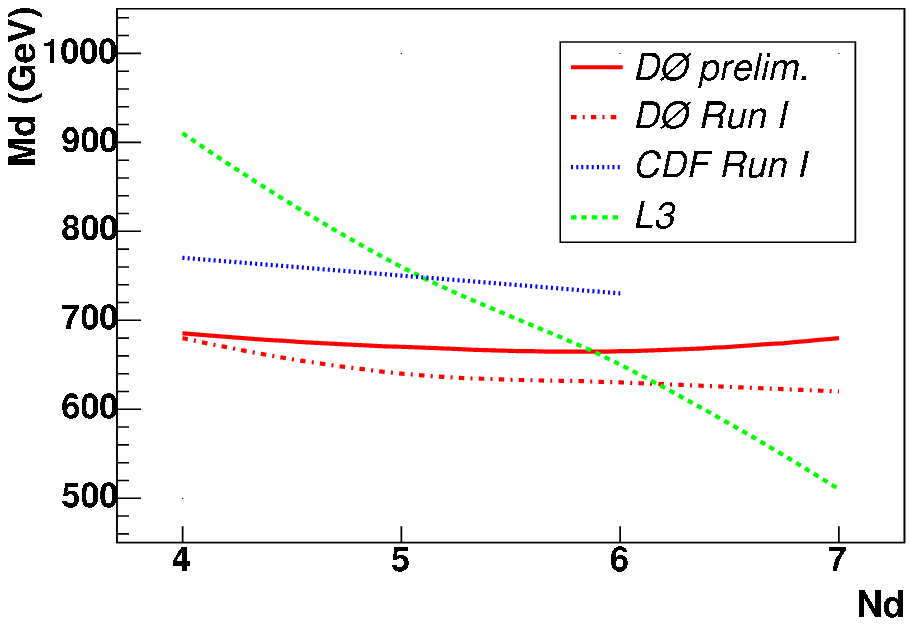}
  \end{tabular}
  \caption{Left: D\O\ mass spectrum for two electromagnetic objects
  in the central calorimeters.
  The shaded histogram indicates the instrumental
  background; the solid curve a fit to background + SM; the dashed
  curve adds a $TEV^{-1}$ LED signal for $M_c = .81\ TeV$.
  Right: The   D\O\ lower limit from jets + missing $E_T$
  on the Planck scale $M_D$ for various numbers of
  extra dimensions, together with earlier limits.}
  \label{fig:eejj}
  \end{center}
  \end{figure}

\subsection {LED with Jets}
D\O\ has also begun searching for Large Extra Dimensions with in
the jets + missing $E_T$ channel, with a smaller sample of 85/pb.
Here, a graviton could recoil against a jet and escape the
detector unseen, leading to a monojet-like topology.  The search
requires a leading jet $P_T
> 150$ GeV, a second jet, but with $P_T < 50$ GeV, and missing
$E_T
> 150$ GeV; the azimuthal separation of the jet and missing $E_T$
direction must be greater than 30 degrees.  The analysis sees 63
events, though $100 \pm 6 \pm 7$ are expected, giving a
better-than-expected limit of 84 events. Interpretation is
currently limited by knowledge of the MC and data jet energy
scales, which results in an uncertainty of the efficiency of
$20\%$ and the background of $+50\%$ to $- 30\%$. The result is a
limit as a function of the number of extra dimensions (see Fig 3b)
The resulting limit is better than the D\O\ Run I
result~\cite{d0j}, but not as good as the CDF Run I
result~\cite{cdfj}; the LEP~\cite{lepj} limit depends on the
number of dimensions differently than at the Tevatron.

\section{Z$^{\prime}$ Searches} The ee mass spectrum can also be
searched for enhancements due to possible Z$^{\prime}$
resonances~\cite{zprime}; both experiments had done so on their
respective 200 event/pb samples. The Z$^{\prime}$ is a spin one
object, but its coupling to pairs of light leptons is
model-dependent. Assuming Standard Model couplings would give a
large cross section contribution, so that such a Z$^{\prime}$
would be relatively easily detected.  CDF~\cite{cdfee} produced a
lower mass limit on a Z$^{\prime}$ with SM couplings of $690$ GeV
from Run I data. They obtain a limit of $750$ GeV from a run II
analysis using the $\sigma \cdot B$ limits based on spin 1
acceptance as a function of boson mass, compared with the
expected values with the SM couplings.  D\O\ produced
limits~\cite{d0zp} of $670$ and $780$ GeV for Run I and II
respectively.  The D\O\ Run II analysis was based on a Pythia
Z$^{\prime}$ simulation for acceptance and cross section
prediction, and a search window optimized for each $M_Z$ to
produce limits on $A \sigma \cdot B$ where $A$ is the acceptance;
the limits were set in terms of ratios of Z$^{\prime}$ to Z cross
section to minimize systematic errors.

Each group also considered the weaker couplings implied by an E6
GUT~\cite{e6}. For 4 types of Z$^{\prime}$, $Z_I, Z_\chi, Z_\psi,
Z_\eta$, the lower limits found by CDF are 570, 610, 625, and 650
GeV, while D\O\ finds lower limits of 575, 640, 650, and 680 GeV.

The cross section calculations used to derive these limits were
notably different between the two experiments, and this may
explain some of the difference in the limits derived.

\section{Other CDF Limits}
\subsection{Limits from the ee Spectrum for Other Models}
CDF summarizes its ee mass spectrum results as a limit on a
$\sigma \cdot B$ excess (beyond the standard model)as a function
of mass for three spins: spin 0, spin 1, and spin 2. These limits
are in the range of .05 to .2 pb .   The three mass limit curves
then may be compared with $\sigma \cdot B$ curves from various
models with the appropriate spin of a particle decaying to an ee
final state. Where the models (as a function of parameters) cross
the limits gives the minimum allowable mass parameter of the
model. Since this requires only the model cross section to be
calculated (the efficiency having been pre-calculated based on
spin and mass), this process is relatively simple, and many
models can be examined rapidly (though not in a fully-optimized
way). Figure 4 shows the results for the Little Higgs model.
Indicative limits for other models~\cite{cdfee} are R Parity
violating sneutrino mass $>$ 630 GeV for $(\lambda^{\prime})^2 B
= .01$ , and Randall-Sundrum graviton mass $>$ 500 GeV for
$k/M_{Pl} = .05$.

\begin{figure}[ht]
  \begin{center}
  \begin{tabular}{cc}
  \includegraphics[width=77mm,height=2.5in]{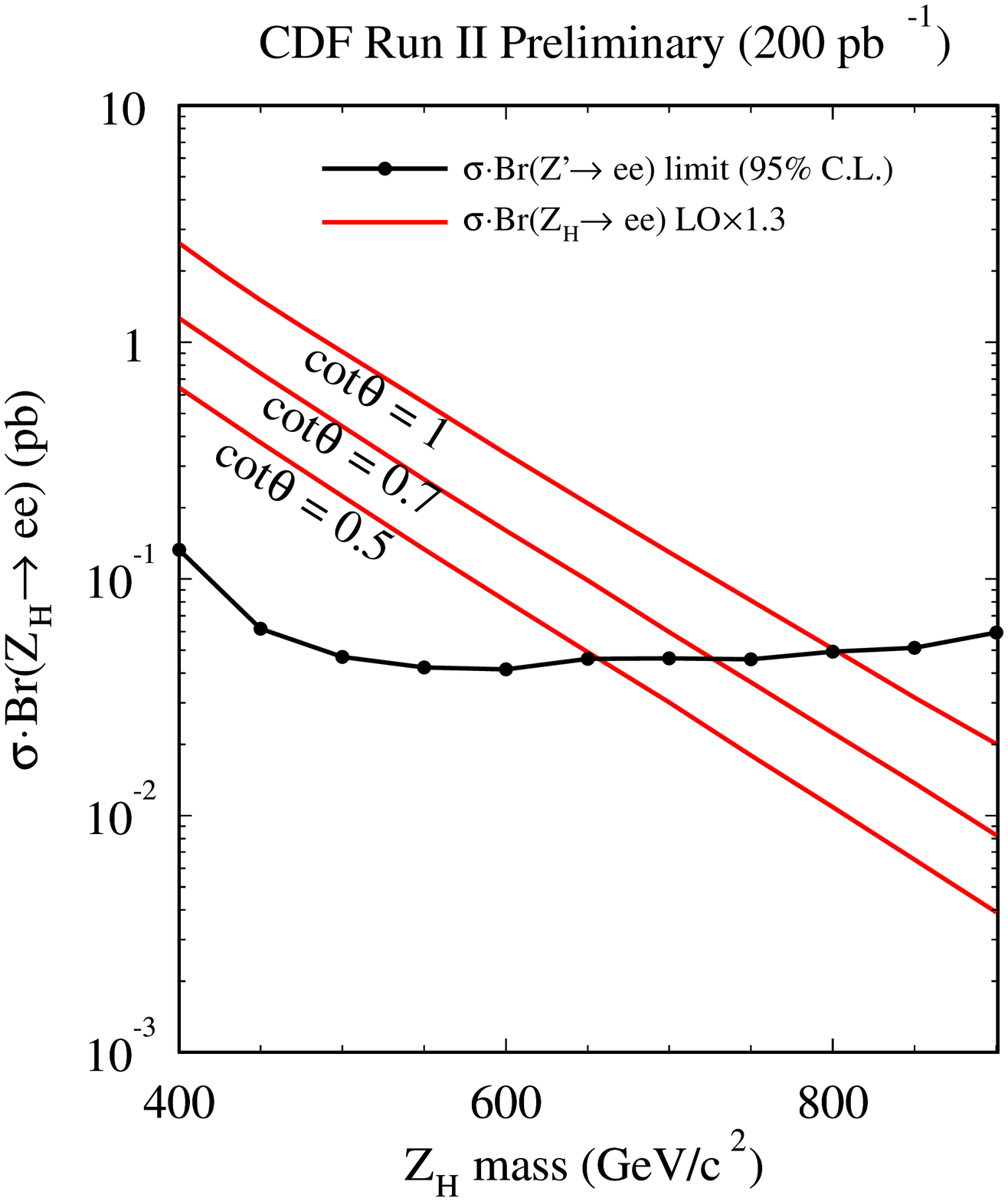} &
    \includegraphics[width=77mm,height=2.5in]{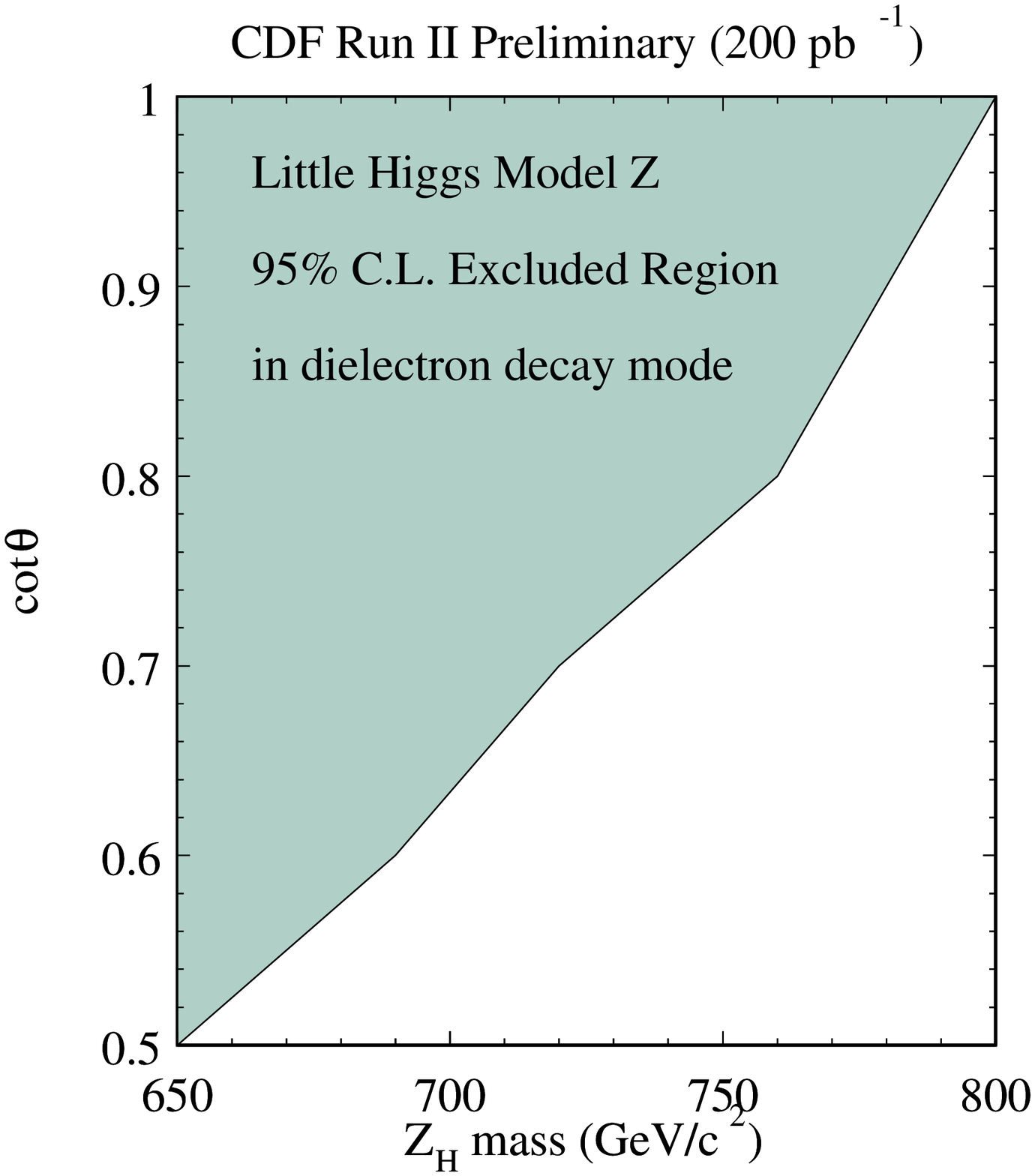}
  \end{tabular}
  \caption{Left: CDF Spin 1 excluded $\sigma \cdot B$  vs. mass
  for ee and
  predicted cross section for various little Higgs models.
  Right: The resulting excluded little Higgs parameters
  as a function of mass.}
  \label{fig:cdflhiggs}
  \end{center}
  \end{figure}

\subsection{CDF Forward ee Events}
CDF has begun exploration of an event sample~\cite{cdffwd} with
two electrons in its forward  calorimetry. This sample has higher
QCD backgrounds than samples requiring one electron in the
central calorimetry. Based on a 173/pb sample, there are some
hints of a possible excess over the SM + backgrounds at large
masses. The observations (backgrounds) for M $\geq$ 250, 300, 350
and 500 are  10 ($5.0 \pm 1.2$), 8 ($2.5 \pm 0.7$) , 3 ($1.4 \pm
.3$), and 2 ($.21 \pm .04$). D\O\ sees 1 event with mass beyond
450, with $.9 \pm .2$ expected background.

\subsection{Time of Flight}
In addition, one other non-ee analysis was derived from Time of
Flight (TOF) analysis~\cite{tof} looking for charged massive
stable particles.  For example, if a stop quark happened to be
sufficiently stable to hit the TOF counters, it would have to have
a mass greater than 95 GeV.

\section{D\O\ SUSY Searches}
The remainder of the analyses discussed here are searches for
production and decay of supersymmetric particles.  At time of this
conference, only D\O\ presented updated SUSY searches. The
analyses and their selection criteria are described
elsewhere~\cite{d0page} in more detail than given here.

\subsection{Squark and Gluino Search}
In R-parity conserving Supersymmetry~\cite{haber}, associated
production of squarks and gluinos could result in pairs of stable
lightest supersymmetric particles (LSPs) escaping undetected.  A
squark would decay in to a quark and a LSP, while a gluino would
decay into a q qbar pair (jets) and another LSP.  Thus one expects
events with a-coplanar jets and missing $E_T$.  The
search~\cite{sqgl} requires two jets, above 50 and 60 GeV $P_T$,
again with a 30 degree azimuthal ($\phi$) separation between the
jet and missing $E_T$ directions to avoid mis-measured jets
simulating real missing $E_T$, and the two jets less than 165
degrees apart in $\phi$.  The final cuts were the minimum missing
$E_T$ and $H_T$ (the scalar $P_T$ sum of the jets). The optimum
cuts values were found by minimizing (with Monte Carlo data) the
expected limit cross section, under the assumption of only
standard model processes entering the Monte Carlo sample. Notice
that this optimization trades off characteristics of the
background and its uncertainty, but does not directly optimize
detection of any particular signal model.  The chosen minimum
values are 175 GeV (missing $E_T$) and 275 GeV ($H_T$).

\begin{figure}[ht]
  \begin{center}
  \begin{tabular}{cc}
  \includegraphics[width=77mm]{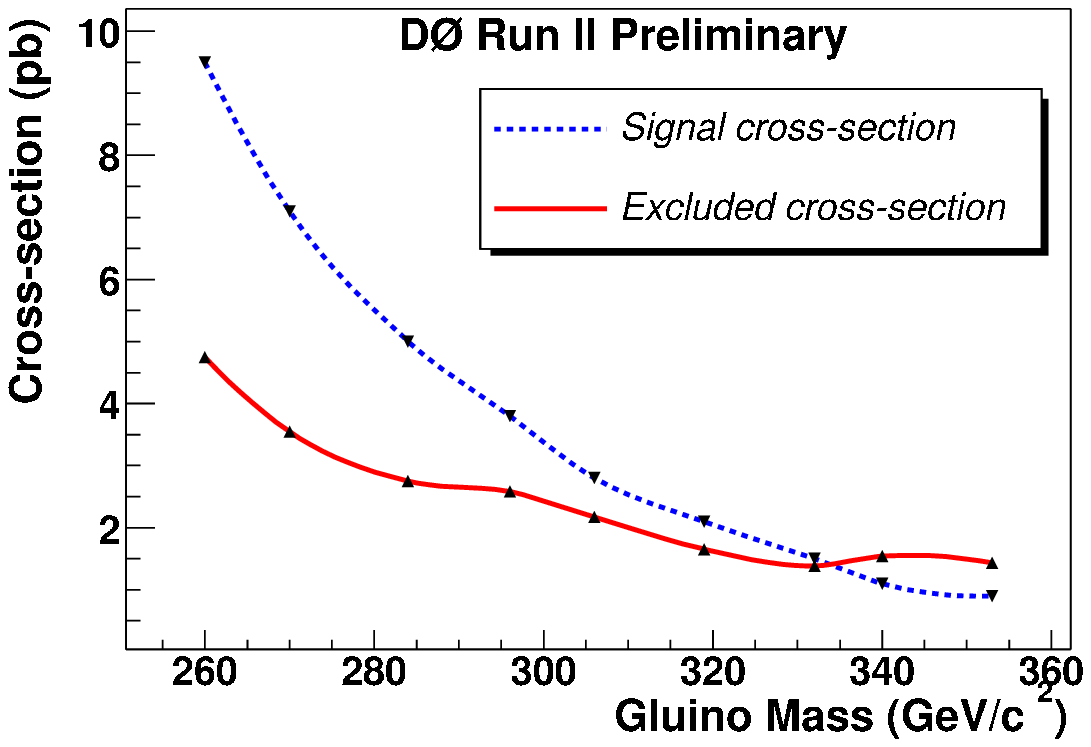} &
    \includegraphics[width=77mm]{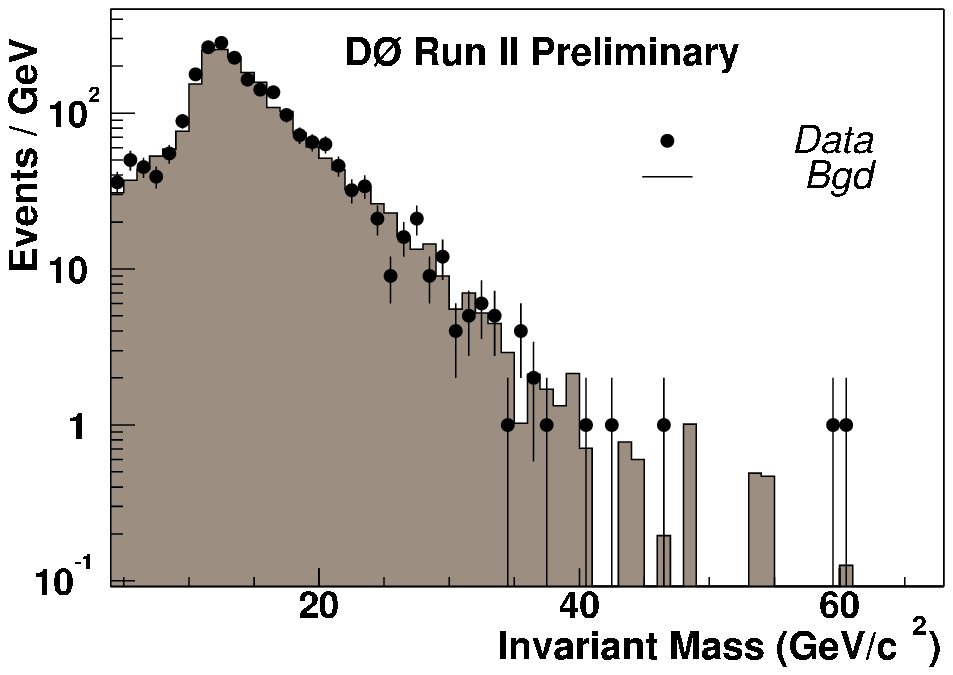}
  \end{tabular}
  \caption{Left: D\O\ cross section limits and predicted
  gluino cross sections.
  Right: D\O\ isolated like-sign dimuon data and the estimated
  background due to b and c decays scaled from the like-sign
  nearly-isolated dimuon sample.}
  \label{fig:susy1}
  \end{center}
  \end{figure}

With these selection criteria, 4 events are found with $2.7 \pm
1.0$ expected.  SUSY efficiencies are of order 5\% for some
typical parameters chosen, though these cuts are more optimal for
light squarks than for light gluinos.  Based on a preliminary
scan of $M_{1/2}$ for $M_o$ = 25, $A_o$ = 0, tan$\beta$ = 3, and
$\mu < 0$, lower limits were found of 290 (squark mass) and 333
GeV (gluino mass); the latter is shown in Fig 5a.  These limits
are already better than Run I limits, despite the smaller sample,
because of the superior QCD background rejection
of
the upgraded D\O\ detector.

\subsection{Trilepton Searches}
One promising search strategy for SUSY at the Tevatron seeks
decays  into a trilepton final state, again with escaping LSP's
providing missing $E_T$; for this final state, standard model
backgrounds are low.  In mSUGRA, such decays may be important for
favorable mass relations among charginos and neutralinos. When
$m_{\chi_1^{\pm}} \approx m_{\chi_2^{0}} \approx 2
m_{\chi_1^{0}}$, leptonic branching fractions are enhanced by low
slepton masses. Still, such a search is challenging as $\sigma
\cdot  B$ is below a picobarn, and there may be soft decay
leptons which are difficult to detect.  The strategy in the D\O\
search is to choose cuts for individual channels which leave
little standard model background, and then to combine the ee,
$\mu \mu$ and e$\mu$ channel results to produce a final limit.

\subsubsection{Like Sign Dimuons}
This analysis~\cite{lsmu} uses a data sample of 147 events/pb.
The main cuts require two like-sign isolated muons of at least 5
and 11 GeV $P_T$, with a missing $E_T$ of at least 15 GeV. QCD
and b and c decays are further suppressed by requiring the $\phi$
difference between the two muons to be less than 2.7 radians (far
from back to back). The remaining b background is scaled from
like-sign muons one of which fails isolation criteria.  The
scaling was tested by estimating the b background in
opposite-sign dimuons. The result of scaling nearly-isolated
muons to estimate the remaining backgrounds in the like sign
sample is shown in Fig 5b; the data is very close to the
estimated background shape. The cuts were chosen before looking at
the final data sample; they were more aimed at reducing
backgrounds than optimizing for any particular SUSY parameters.
The analysis finds 1 event with an expected background of .13
events. For several SUSY points examined, .2 to .4 SUSY events
are expected to pass the cuts for the sample size. and $\sigma
\cdot B \sim .3 $pb.

\subsubsection{e$\mu$}
This analysis~\cite{emu} was performed on a 158 event/pb sample.
An electron of 12 GeV and an isolated muon of 8 GeV were sought in
an event with between 15 and 80 GeV of missing $E_T$.  A number
of other kinematic criteria were used to suppress WW and W+jet
backgrounds, including a requirement on the vector sum of the
detected leptons and missing $E_T$ be less than 6 GeV. Figure 6a
shows the data and background contributions before this last cut.
After this cut, these preliminary criteria left a standard model
background of $2.9 \pm .4$ events, while one event was observed.
One can also add an explicit but loose requirement for a third
lepton by demanding another isolated track of greater than 3 GeV
$P_T$. This reduces the expected background to $0.5 \pm .2$
events, while no such events were seen.  The third lepton
requirement loses little efficiency for SUSY events.  For typical
SUSY parameters,  .6 to .9 events were expected.

\begin{figure}[ht]
  \begin{center}
  \begin{tabular}{cc}
  \includegraphics[width=77mm,height=2.5in]{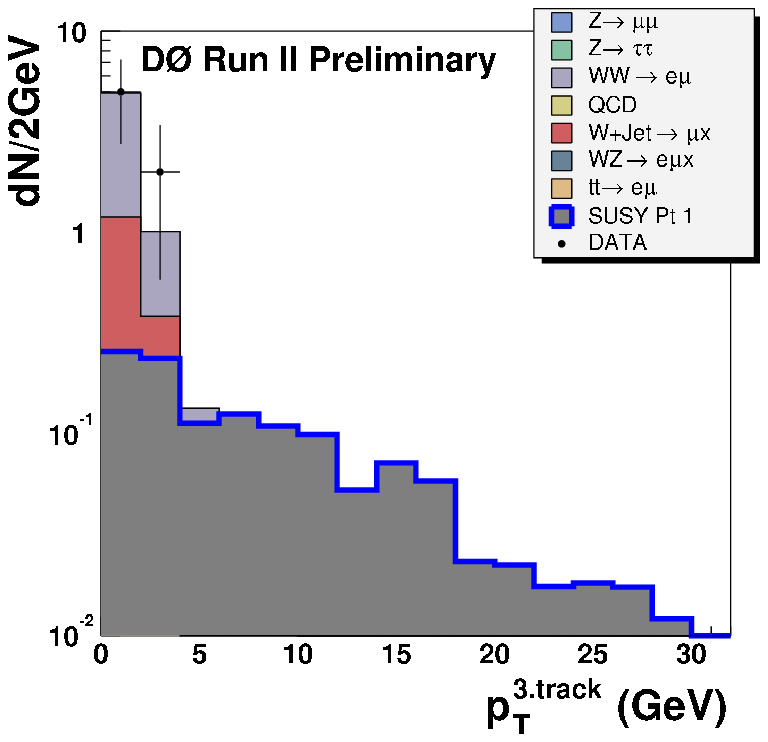} &
    \includegraphics[width=77mm]{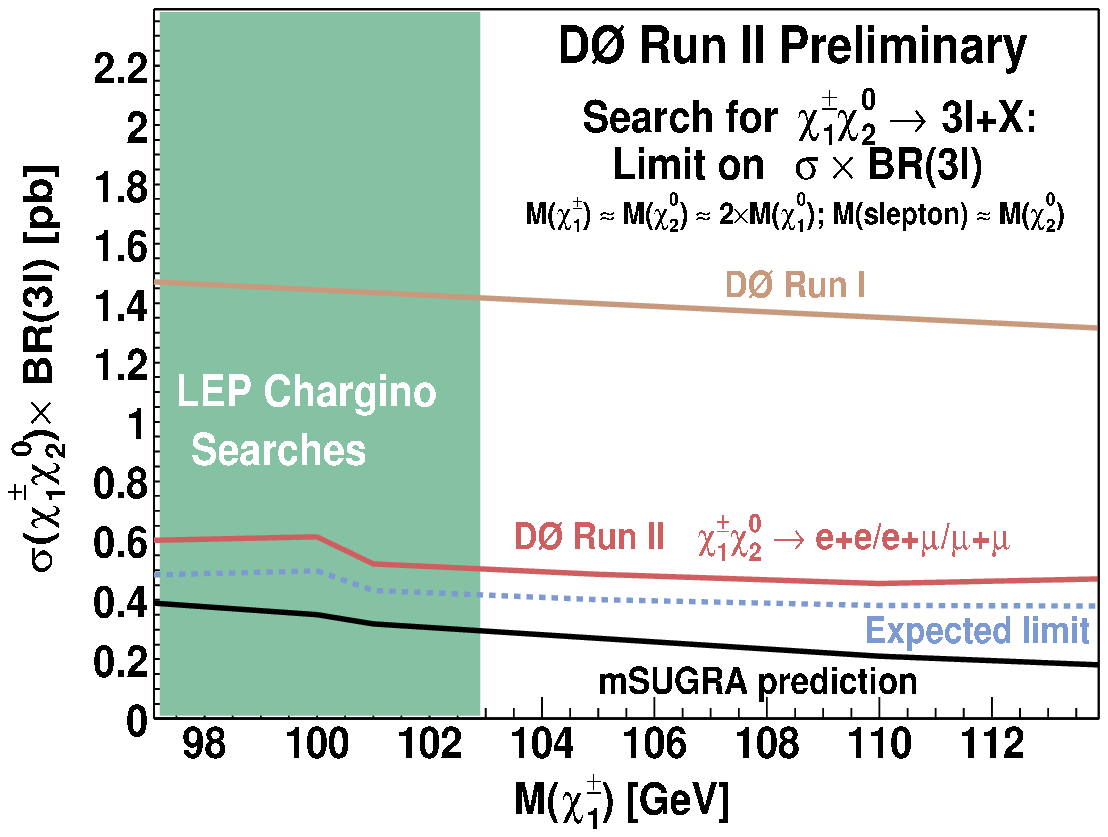}
  \end{tabular}
  \caption{Left: The data and the contributions of various
  backgrounds to the D\O\
  $e\mu$ SUSY analysis before requiring a third hard track.
  Right: The limit in $\sigma \cdot B$ for trileptons from the D\O\
  analysis for Run I and Run II.  The Run II limit does not yet
  exclude the mSUGRA models examined.}
  \label{fig:susy2}
  \end{center}
  \end{figure}

\subsubsection {ee}
The final dilepton analysis~\cite{susyee} considered a  174
events/pb sample. It required events with missing $E_T > 20$ GeV
and two electrons above 12 and 8 GeV $P_T$, one from the central
rapidity region. A third lepton was required by asking for 3
tracks above 3 GeV, with one well separated from the electrons.
No requirement was made on the sign of tracks.  The W, Z and top
backgrounds were suppressed by various kinematic cuts, including
no jets above 80 GeV, separation in $\phi$ of the electrons less
than 2.8 radians, ee mass $< 60$, and transverse mass $> 15$ GeV,
separation of the electron and the missing $E_T$ direction, and a
product $\pi= $ track $P_T\ \times $ missing $E_T > 250\ GeV^2$.
After all these cuts, an expected standard model background of
$.3 \pm .4$ remained, and one event was observed.  Signal
expectations for typical models ranged from .8 to 1.6 events.

\subsection {Combined Trilepton Results}
The three channels were combined~\cite{combined} with the $CL_S$
technique~\cite{cls}.  The inputs were the results of the $\mu\mu$
analysis,  three independent bins of the $\pi$ parameter from the
ee channel, and  the two final subsets of the $e\mu$ analysis.



The results of the combined searches is show in Fig 6b. Plotted
as a function of the lowest chargino mass is the cross section
for joint production of this chargino with the 2nd lightest
neutralino, multiplied by the branching ratio to trileptons.  The
curve shows a substantial improvement over run I, and $\sigma
\cdot B$ limits less than .5 pb, just a bit above the expected
limit based on no physics beyond the standard model. However, the
exclusion contour is still not sufficiently sensitive to rule out
the expected production rate for the chosen SUSY points.  The
present scan is over the region $M_o, M_{1/2}$ (72, 165) to (88,
185), with $A_o = 0$, tan $\beta = 3$, $\mu < 0$, which covers a
region near and beyond the LEP II chargino mass limit of some 103
GeV for large slepton/neutrino masses~\cite{lepino}.

\section{Summary}
Unfortunately, nature has declined to flood us with new phenomena
readily visible at the Tevatron--at least so far.  Clear
improvements over Run I limits are already visible.  While the
sample is roughly twice that of Run I, and the center of mass
energy is somewhat higher, the fact that the limits are already
improved over Run I does imply that the detectors are
fundamentally working and that the understanding of the upgraded
detectors is making good progress.  Although all analyses are not
fully optimized, many are already producing limits for $\sigma
\cdot B$ of  .5 pb or better.  That means we are talking about
sensitivity to 1-10 events for processes with 1-10\% efficiency.
Clearly publication on these topics is in our near future.

\section*{Acknowledgements}
It is a pleasure to thank my many colleagues, without whom this
report would have been impossible, for sharing their expertise,
effort and enthusiasm.  The author  acknowledges support from the
National Science Foundation under Grants PHY-0140106 and 0334296.
Any opinions, findings, and conclusions or recommendations
expressed in this material are those of the author(s) and do not
necessarily reflect the views of the National Science Foundation.

\end{document}